# Anisotropic Magnetoresistance Effect: General Expression of AMR Ratio and Intuitive Explanation for Sign of AMR Ratio

Satoshi Kokado[1, a] and Masakiyo Tsunoda[2]

[1]Graduate School of Engineering, Shizuoka University, Hamamatsu 432-8561, Japan

[2]Graduate School of Engineering, Tohoku University, Sendai 980-8579, Japan

[a]tskokad@ipc.shizuoka.ac.jp

**Abstract.** We derive the general expression of the anisotropic magnetoresistance (AMR) ratio of ferromagnets for a relative angle between the magnetization direction and the current direction. We here use the two-current model for a system consisting of a spin-polarized conduction state (*s*) and localized d states (*d*) with spin-orbit interaction. Using the expression, we analyze the AMR ratios of Ni and a half-metallic ferromagnet. These results correspond well to the respective experimental results. In addition, we give an intuitive explanation about a relation between the sign of the AMR ratio and the *s-d* scattering process.

**Introduction**

The anisotropic magnetoresistance (AMR) effect, in which the electrical resistivity depends on a relative angle $\theta$ between the magnetization ($M_{ex}$) direction and the electric current ($I$) direction, has been studied extensively both experimentally [1-5] and theoretically [1,6]. The AMR ratio is often defined by

$$\frac{\Delta\rho(\theta)}{\rho} = \frac{\rho(\theta) - \rho_\perp}{\rho_\perp}, \tag{1}$$

with $\rho_\perp = \rho(\pi/2)$, where $\rho(\theta)$ is a resistivity for $\theta$. Theoretically, Campbell-Fert-Jaoul (CFJ) [1] derived an expression of $\Delta\rho(0)/\rho$ of strong ferromagnets such as Ni-based alloys. Recently, we have extended the CFJ model to a more general theory [6] which is effective in examining $\Delta\rho(0)/\rho$'s of the strong, weak, and half-metallic ferromagnets. This theory is based on the two-current model for a system consisting of a spin-polarized conduction state (*s*) and localized d orbitals (*d*) with spin-orbit interaction, where the conduction state contains the s, p, and conductive d states. We have analyzed $\Delta\rho(0)/\rho$'s of various ferromagnets using the respective approximate expressions [6,3].

For this theory, however, we did not perform the following two tasks: (i) derivation of the general expression of $\Delta\rho(\theta)/\rho$ and its applications and (ii) determining an intuitive explanation for a relation between the sign of $\Delta\rho(0)/\rho$ and the *s-d* scattering process, where the *s-d* scattering represents that the conduction electron is scattered into the localized d orbitals by nonmagnetic impurities. Such studies appear to play an important role in the understanding and analyses of the AMR effect. In this paper, we address the above tasks. In particular, we investigate the AMR effects of Ni and the half-metallic ferromagnet (HMF).

**Model and Method**

To obtain $\Delta\rho(\theta)/\rho$, we extend our theory to a model with the angle $\theta$ between the $\boldsymbol{M}_{\text{ex}}$ direction and the $\boldsymbol{I}$ direction. In this model, resistivity of the $\sigma$ spin ($\sigma = \uparrow$ or $\downarrow$), $\rho_\sigma(\theta)$, is expressed as

$$\rho_\sigma(\theta) = \rho_{s\sigma} + \sum_{\varsigma=\uparrow,\downarrow} \rho_{s\sigma \to d\varsigma}(\theta), \qquad (2)$$

with $\rho_{s\sigma} = m_\sigma^*/(n_\sigma e^2 \tau_{s\sigma})$, $\rho_{s\sigma \to d\varsigma}(\theta) = \sum_{M=0,\pm1,\pm2} m_\sigma^* / \left[ n_\sigma e^2 \tau_{s\sigma \to dM\varsigma}(\theta) \right]$, $1/\tau_{s\sigma} \propto |V_s|^2 D_{s\sigma}$, and $1/\tau_{s\sigma \to dM\varsigma}(\theta) \propto \left| \langle dM\varsigma | V_{\text{imp}} | s\sigma(\theta) \rangle \right|^2 D_{d\varsigma}$. Here, $n_\sigma$ ($m_\sigma^*$) is the number density (effective mass) of the electrons in the conduction band of the $\sigma$ spin and $e$ is the electronic charge. The quantity $V_{\text{imp}}$ is a nonmagnetic impurity potential defined by a screened Coulomb potential [7], and $V_s$ is the matrix element of $V_{\text{imp}}$ between the conduction states. The state $|s(\theta)\rangle$ is the conduction state of the $\sigma$ spin represented by the plane wave. The orbital $|dM\varsigma\rangle$ ($M=0, \pm1, \pm2$ and $\varsigma = \uparrow$ or $\downarrow$) is the localized d orbital, given by

$$|dM\varsigma\rangle = \left[1 - f_{M\varsigma}(\lambda)\right] |\varphi_{M\varsigma}\rangle + \sum_{m=0,\pm1,\pm2\,(\neq M)} \sum_{\sigma=\uparrow,\downarrow\,(\neq \varsigma)} g_{m\sigma}^{M\varsigma}(\lambda) |\varphi_{m\sigma}\rangle, \qquad (3)$$

with $f_{M\varsigma}(0) = g_{m\sigma}^{M\varsigma}(0) = 0$, and $0 < f_{M\varsigma}(\lambda) \ll 1$ and $|g_{m\sigma}^{M\varsigma}(\lambda)| \ll 1$ for $\lambda \neq 0$, where $\lambda$ is the spin-orbit coupling [6]. Here, $\left[1 - f_{M\varsigma}(\lambda)\right] |\varphi_{M\varsigma}\rangle$ is the dominant orbital, while $\sum_{m(\neq M)} \sum_{\sigma(\neq \varsigma)} g_{m\sigma}^{M\varsigma}(\lambda) |\varphi_{m\sigma}\rangle$ is the hybridized orbital due to the spin-orbit interaction, where $\varphi_{M\varsigma} = \varphi_M \chi_\varsigma$, $\varphi_0 \propto (3z^2 - r^2)/(2\sqrt{3})$, $\varphi_{\pm1} \propto \mp z(x \pm iy)/\sqrt{2}$, and $\varphi_{\pm2} \propto (x \pm iy)^2/(2\sqrt{2})$, with $\chi_\varsigma$ being the spin state and $(x, y, z)$ being the position of the electron. The quantities $M$ and $\varsigma$ are the magnetic quantum number and the spin of the dominant orbital, respectively. The coefficients $f_{M\varsigma}(\lambda)$ and $g_{m\sigma}^{M\varsigma}(\lambda)$ indicate the strength of the hybridization. This $|dM\varsigma\rangle$ is actually obtained by applying a perturbation theory to the single atom Hamiltonian, which consists of the unperturbed term of a hydrogen-like atom with Zeeman interaction due to the exchange field $\boldsymbol{H}_{\text{ex}}$ ($\propto \boldsymbol{M}_{\text{ex}}$) and the perturbed term of the spin-orbit interaction. In addition, $D_{s\sigma}$ ($D_{d\varsigma}$) is the density-of-states (DOS) of the conduction state of the $\sigma$ spin at the Fermi energy, $E_{\text{F}}$ (that of the localized d orbital of the $\varsigma$ spin at $E_{\text{F}}$).

**Results and Consideration**

In the same manner as our previous study [6], we first obtain $\rho_\sigma(\theta)$ as

$$\rho_\uparrow(\theta) = \rho_{s\uparrow} + \frac{3}{2}\gamma \rho_{s\uparrow \to d\downarrow} + \left(1 - \frac{3}{2}\gamma\right) \rho_{s\uparrow \to d\uparrow} + \frac{\gamma}{2}\left(\rho_{s\uparrow \to d\downarrow} - \rho_{s\uparrow \to d\uparrow}\right) \cos(2\theta), \qquad (4)$$

$$\rho_\downarrow(\theta) = \rho_{s\downarrow} + \left(1 - \frac{3}{2}\gamma\right) \rho_{s\downarrow \to d\downarrow} + \frac{3}{2}\gamma \rho_{s\downarrow \to d\uparrow} + \frac{\gamma}{2}\left(\rho_{s\downarrow \to d\uparrow} - \rho_{s\downarrow \to d\downarrow}\right) \cos(2\theta), \qquad (5)$$

with $\rho_{s\sigma \to d\varsigma} = m_\sigma^*/(n_\sigma e^2 \tau_{s\sigma \to d\varsigma})$ [6], $\gamma = (3/4)(\lambda/H_{\text{ex}})^2$, and $H_{\text{ex}} = |\boldsymbol{H}_{\text{ex}}|$. Substituting Eqs. (4) and (5) into an expression of total resistivity including the spin-flip scattering (see Eq. (7) in Ref. [6]), we derive the general expression of $\Delta\rho(\theta)/\rho$, i.e.,

$$\frac{\Delta \rho(\theta)}{\rho} = AMR \frac{1}{2}\left[1 + \cos(2\theta)\right], \quad (6)$$

with $\Delta\rho(0)/\rho = AMR$ and $AMR = (A+B)/(CD)$, where A, B, C, and D are the respective ones in Eq. (28) of Ref. [6]. When the spin-flip terms in AMR are ignored, AMR eventually becomes

$$AMR = \frac{-\gamma}{(\beta_\uparrow y_\uparrow)^{-1}+1}(1-x_d)\left\{\frac{\frac{r_{m*}^4}{x_s^2} - \frac{\beta_\downarrow}{\beta_\uparrow x_s}\left[\frac{1+\beta_\uparrow y_\uparrow}{1+(x_d/x_s)\beta_\downarrow y_\uparrow}\right]^2}{\frac{r_{m*}^4}{x_s^2} + \frac{1+\beta_\uparrow y_\uparrow}{1+(x_d/x_s)\beta_\downarrow y_\uparrow}}\right\}, \quad (7)$$

with $x_s = D_{s\downarrow}/D_{s\uparrow}$, $x_d = D_{d\downarrow}/D_{d\uparrow}$, $y_\sigma = D_{d\sigma}/D_{s\sigma}$, $\beta_\sigma = N_n |V_{s\sigma \to d\sigma}|^2/|V_s|^2$, and $r_{m*} = m_\downarrow^*/m_\uparrow^*$, where $V_{s\sigma \to d\sigma}$ is the matrix element of $V_{imp}$ between the conduction state and the localized d orbital and $N_n$ is the number of nearest-neighbor atoms around a single impurity [6]. We note here that the $\cos(2\theta)$ terms in Eqs. (4) - (6) are related to the two-fold rotational symmetry of $\varphi_0$ and $\varphi_{\pm 2}$ [8]. Furthermore, $\cos(2\theta)$ in Eq. (6) is the dominant component in the $\theta$ dependence of the AMR ratio observed in the experiments [5].

Using Eq. (7) (i.e., $\Delta\rho(0)/\rho$), we evaluate AMR's of Ni and HMF. Here, the dominant s-d scattering processes $s\sigma \to d\varsigma$ of Ni and HMF are $s\uparrow \to d\downarrow$ and $s\uparrow \to d\uparrow$, respectively [6]. The notation $s\sigma \to d\varsigma$ means that the conduction electron of the $\sigma$ spin is scattered into "the $\sigma$ spin state in the localized d orbital of M and $\varsigma$ at $E_F$" by impurities. The partial DOS's for the dominant and hybridized d orbitals [6] are shown in Fig. 1(c). In this calculation, $\gamma = 0.01$ [6], $r_{m*} = 1$ and $\beta_\uparrow = \beta_\downarrow = \beta$ are assumed. In addition, ($x_s$, $x_d$, $\rho_{s\downarrow}/\rho_{s\uparrow}$) of Ni is set to be (0.32, 10, 10) on the basis of the theoretical data [6,9], where $x_s$ has been estimated from $x_s = r_{m*}^2/\sqrt{\rho_{s\downarrow}/\rho_{s\uparrow}}$ of Eq. (48) in Ref. [6]. As for HMF, Eq. (7) is rewritten as $AMR = -\gamma/[(\beta y_\uparrow)^{-1}+1]$ on the assumption of $x_s = x_d = x_D$ and $x_D \to 0$ or $\infty$. Figure 1(a) shows the $\beta y_\uparrow$ dependence of their AMR's. The sign of AMR of Ni is positive, while that of HMF is negative. These signs agree with the respective ones observed in experiments [2-4]. Furthermore, when $\beta y_\uparrow$'s of Ni and HMF are, respectively, chosen to be $\beta y_\uparrow \sim 0.3$ and 0.7, AMR's correspond to the respective experimental values, i.e., 2 % for Ni [2] and $-0.4$ % for HMF [3].

Let us consider the relation between the sign of AMR and the dominant $s\sigma \to d\varsigma$ scattering, i.e., "$AMR < 0$ for $s\uparrow \to d\uparrow$" of HMF and "$AMR > 0$ for $s\uparrow \to d\downarrow$" of Ni [6]. The relation is explained by considering that (i) the final state of the scattering is the component of the ***I*** direction of the distorted d orbitals, and (ii) the distorted d orbitals are oriented by ***M***$_{ex}$ through the spin-orbit coupling (see Figs.1 (b) and (c)).

(i) The selection rule for the s-d scattering requires that the final state is $|\varphi_{0\sigma}\rangle$ in the case of $\theta = 0$, while that is $|\varphi_{\pm 2\sigma}\rangle$ and $|\varphi_{0\sigma}\rangle$ in the case of $\theta = \pi/2$, where the spin of the conduction electron is conserved [1,6]. In short, the final states roughly correspond to "the states that exist along the axis of the ***I*** direction". These states are shown by the blue and red bars in ellipses (see (ii), below). In addition, $\rho_{s\sigma \to d\varsigma}(\theta)$ increases with increasing the magnitude of probability amplitude of such states (i.e., the length of the bars).

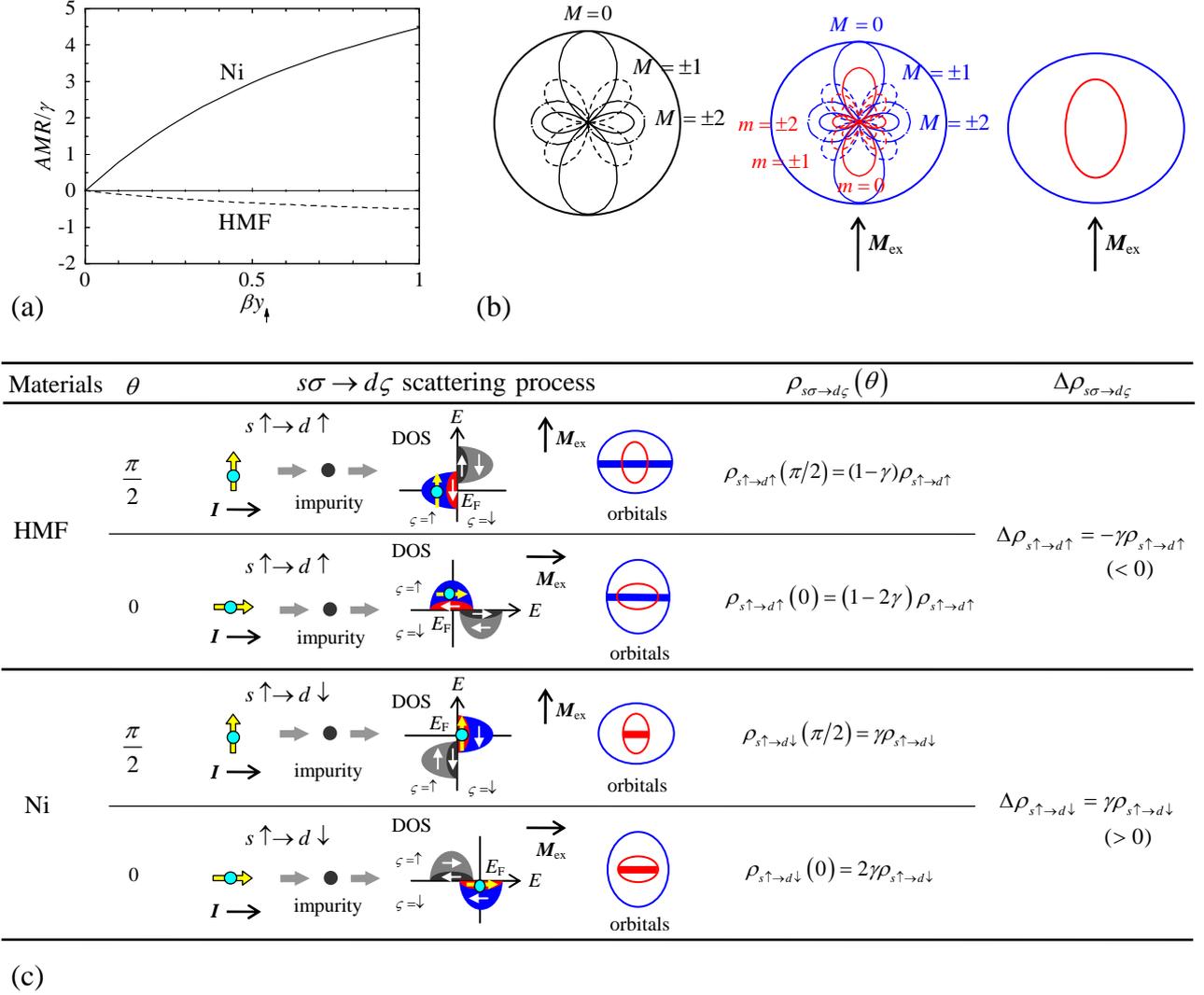

Figure 1 (a) $\beta y_\uparrow$ dependence of *AMR*'s of HMF and Ni. Here, *AMR*'s of HMF and Ni are shown by the dashed and solid curves, respectively. (b) Left panel: Pure d orbitals ($\lambda = 0$) enclosed by a circle. The d orbitals are shown by the direction dependence of the magnitude of probability amplitude. Middle panel: Distorted d orbitals ($\lambda \neq 0$) enclosed by an ellipse. The blue and red orbitals are the dominant and hybridized orbitals, respectively, where both have mutually opposite spins. The d orbitals are oriented by $M_{ex}$. Note that the distortion and hybridization effects have been exaggeratedly illustrated for clarity. Right panel: Schematic illustration of the middle panel. (c) Dominant $s\sigma \to d\varsigma$ scattering process, $\rho_{s\sigma \to d\varsigma}(0)$, $\rho_{s\sigma \to d\varsigma}(\pi/2)$, and $\Delta\rho_{s\sigma \to d\varsigma} = \rho_{s\sigma \to d\varsigma}(0) - \rho_{s\sigma \to d\varsigma}(\pi/2)$ for HMF or Ni. The *d* part shows the d orbitals and the d-band partial DOS's [6]. The blue and red DOS's are the partial DOS's for the dominant and hybridized orbitals, respectively.

(ii) The pure d orbital $|\varphi_{M\varsigma}\rangle$ with $\lambda = 0$ is schematically illustrated by a circle because of $\Delta\rho(\theta)/\rho = 0$. Then, $[1 - f_{M\varsigma}(\lambda)]|\varphi_{M\varsigma}\rangle$ of Eq. (3) is regarded as a large ellipse (blue) whose minor axis is oriented along the quantization axis (//$M_{ex}$). In addition, $\sum_{m(\neq M)}\sum_{\sigma(\neq \varsigma)} g_{m\sigma}^{M\varsigma}(\lambda)|\varphi_{m\sigma}\rangle$ of Eq. (3) is a small ellipse (red) whose major axis is oriented along the quantization axis. The distortion from the circle reflects that $f_{0\varsigma}(\lambda)$ ($g_{0\sigma}^{M\varsigma}(\lambda)$) is relatively larger than the other $f_{M\varsigma}(\lambda)$'s (the other $g_{m\sigma}^{M\varsigma}(\lambda)$'s) [6].

On the basis of the above (i) and (ii), we obtain $\rho_{s\uparrow \to d\uparrow}(0)$ and $\rho_{s\uparrow \to d\uparrow}(\pi/2)$ of HMF as $(1-2\gamma)\rho_{s\uparrow \to d\uparrow}$ and $(1-\gamma)\rho_{s\uparrow \to d\uparrow}$, respectively. In addition, $\rho_{s\uparrow \to d\downarrow}(0)$ and $\rho_{s\uparrow \to d\downarrow}(\pi/2)$ of Ni are $2\gamma\rho_{s\uparrow \to d\downarrow}$ and $\gamma\rho_{s\uparrow \to d\downarrow}$, respectively. When the numerator of $\Delta\rho(0)/\rho$ (see Eq. (1)) is approximated as $\Delta\rho_{s\sigma \to d\varsigma} = \rho_{s\sigma \to d\varsigma}(0) - \rho_{s\sigma \to d\varsigma}(\pi/2)$ by focusing on only the dominant $s\sigma \to d\varsigma$ scattering [10], $\Delta\rho_{s\sigma \to d\varsigma}$'s of HMF and Ni become $\Delta\rho_{s\uparrow \to d\uparrow} = -\gamma\rho_{s\uparrow \to d\uparrow} < 0$ and $\Delta\rho_{s\uparrow \to d\downarrow} = \gamma\rho_{s\uparrow \to d\downarrow} > 0$, respectively.

**Conclusions**

We derived the general expression of $\Delta\rho(\theta)/\rho$ for the relative angle $\theta$ between the $\boldsymbol{M}_{\rm ex}$ direction and the $\boldsymbol{I}$ direction. The sign and the magnitude of $\Delta\rho(0)/\rho$'s of Ni and HMF corresponded well to the respective experimental results. The relation between the sign of $\Delta\rho(0)/\rho$ and the s-d scattering process was explained by considering that the final state of the scattering was the component of the $\boldsymbol{I}$ direction of the distorted d orbitals, where the d orbitals were oriented by $\boldsymbol{M}_{\rm ex}$.

**Acknowledgements**

This work has been supported by Grant-in-Aids for Scientific Research (C) (No. 25390055) and (B) (No. 23360130) from the Japan Society for the Promotion of Science.